\documentclass[preprint2]{aastex631}
\usepackage{amsmath}

\begin{document}
\nolinenumbers

\newcommand{\mo}[1]{{\textbf{(Modified) #1}}}
\newcommand{\ad}[1]{{\textbf{(Added) #1}}}

\title{Three-Dimensional Simulations of Type Ia Supernova Remnants I: Effects of a Main-Sequence Companion Star}

\correspondingauthor{Zheng-Wei Liu}
\email{zwliu@ynao.ac.cn}

\author{Jingxiao Luo}
\affiliation{Yunnan Observatories, Chinese Academy of Sciences (CAS), Kunming 650216, P.R. China}
\affiliation{International Centre of Supernovae (ICESUN), Yunnan Key Laboratory of Supernova Research,  Kunming 650216, P.R. China}
\affiliation{University of Chinese Academy of Sciences, Beijing 100049, P.R. China}

\author{Gilles Ferrand}
\affiliation{Department of Physics and Astronomy, 30A Sifton Road, University of Manitoba, Winnipeg, Manitoba, R3T 2N2, Canada}
\affiliation{RIKEN Center for Interdisciplinary Theoretical and Mathematical Sciences (iTHEMS), 2-1 Hirosawa, Wako, Saitama 351-0198, Japan}

\author{Zhengwei Liu}
\affiliation{Yunnan Observatories, Chinese Academy of Sciences (CAS), Kunming 650216, P.R. China}
\affiliation{International Centre of Supernovae (ICESUN), Yunnan Key Laboratory of Supernova Research,  Kunming 650216, P.R. China}

\author{Xuefei Chen}
\affiliation{Yunnan Observatories, Chinese Academy of Sciences (CAS), Kunming 650216, P.R. China}
\affiliation{International Centre of Supernovae (ICESUN), Yunnan Key Laboratory of Supernova Research,  Kunming 650216, P.R. China}
\affiliation{Key Laboratory for the Structure and Evolution of Celestial Objects, CAS, Kunming 650216, P.R. China}

\author{Zhanwen Han}
\affiliation{Yunnan Observatories, Chinese Academy of Sciences (CAS), Kunming 650216, P.R. China}
\affiliation{International Centre of Supernovae (ICESUN), Yunnan Key Laboratory of Supernova Research,  Kunming 650216, P.R. China}
\affiliation{Key Laboratory for the Structure and Evolution of Celestial Objects, CAS, Kunming 650216, P.R. China}



\begin{abstract}

Type Ia supernovae (SNe Ia) serve as one of cosmic standard candles, but their exact progenitor channel is still an open question. 
SNe Ia commonly come from binary star evolution. 
Therefore, one of the major differences among the proposed progenitor channels is whether there is a more-or-less intact companion star remaining at the time of explosion, which causes the SN ejecta to be more asymmetrical.
As the SN ejecta evolved into supernovae remnants (SNR), the imprint formed by the companion interaction may affect the morphology of the SNR.
In addition, the progenitor systems may have experienced different mass transfer histories and therefore led to formation of different circumstellar material (CSM) environments, which may also affect the early evolution of SNR. 
In this study, we use GADGET and RAMSES codes to simulate these physical effects and follow the evolution into early-phases of SNRs.
In our simulations, we consider different ejecta models and track the element distribution.
We compare our simulation with actual observations and conclude that despite some SNRs having morphology resemblance to our simulation results, their highly asymmetric expansion rates are hard to explain by interaction between SN ejecta and a companion star alone.

\end{abstract}



\section{Introduction} \label{sec:intro}

Type Ia supernovae (SNe Ia) are commonly believed to be thermonuclear explosions of massive white dwarfs (WDs) \citep{1960ApJ...132..565H, 1967ApJ...150..115F, 1984ApJ...286..644N}.
In addition to various sub-groups \citep[see][for a review]{2017hsn..book..317T}, many SNe Ia share similar characteristics \citep[e.g., light curve shape and spectral features, see][]{1993AJ....106.2383B, 2000ARA&A..38..191H}. 
Since the advent of Philip's Relation \citep{1993ApJ...413L.105P}, it has been possible to further normalize the light curves of standard SNe Ia, as brighter SNe Ia fade slower and dimmer SNe Ia fade faster.
Therefore, these SNe Ia can be used as standard candles to measure distances over cosmic distances.
SNe Ia served crucial roles in the discovery of Dark Energy and the accelerating expansion of the Universe \citep{1998AJ....116.1009R, 1999ApJ...517..565P, 1998ApJ...507...46S}, as well as the ongoing "Hubble Tension" \citep{2021CQGra..38o3001D, 2022ApJ...934L...7R}.

Despite their great importance, we do not fully understand the progenitors of SNe Ia and the details of their explosions \citep[see][for reviews]{2000ARA&A..38..191H, 2012NewAR..56..122W, 2013FrPhy...8..116H, 2016IJMPD..2530024M, 2018SSRv..214...72R, 2019NewAR..8701535S, 2023RAA....23h2001L, 2025A&ARv..33....1R}.
In particular, two major proposed formation channels for SNe Ia: the single-degenerate channel (SD channel) \citep{1973ApJ...186.1007W, 1984ApJ...286..644N} and the double-degenerate channel \citep[DD channel, see][]{1984ApJ...277..355W, 1984ApJ...284..719I, 2010Natur.463...61P}.
In the SD channel, one compact, degenerate carbon-oxygen WD (CO WD) accretes material from its non-degenerate companion and gains mass.
As the mass of CO WD approaches the Chandrasekhar limit ($\mathrm{M_{ch}}$), a runaway nuclear reaction starts near the center of the WD and its flame engulfs and destroys the entire WD, resulting in a SN Ia \citep{2004MNRAS.350.1301H, 2014MNRAS.438.1762F, 2017hsn..book.1185R, 2016MNRAS.459.1781L, 2017MNRAS.470L..72L}.
In the DD channel, two degenerate WDs initially in a tight orbit gradually spiral inward as their orbit decays due to mechanisms such as gravitational waves \citep{1990ApJ...348..647B, 2011ApJ...737...89D}.
As they eventually merge, SN Ia can be triggered in multiple ways \citep{2000ARA&A..38..191H, 2010Natur.463...61P, 2015ApJ...807..105S}.
If the total mass is close or above $\mathrm{M_{ch}}$, a thermonuclear runaway can be triggered near the center of the merger product \citep[e.g.,][]{2014ApJ...785..105M}.
Other ways that may lead to SNe Ia within the DD channel have also been proposed, such as the $\mathrm{D^6}$ model, where a double-detonation occurs at the stage of dynamic mass transfer \citep{2010ApJ...709L..64G, 2021ApJ...922...68S}.
With better knowledge of SNe Ia formation channels, we may better understand the intrinsic scatter among the SNe Ia population, and improve the accuracy of SNe Ia as standard candles.

Observationally, some methods have been proposed to infer the progenitor system of SNe Ia \citep[see][for a review]{2014ARA&A..52..107M}.
In this paper, we mainly focus on the possibility of differentiating the formation channels of SNe Ia via features in their supernova remnants (SNRs), especially by the dynamic and morphology features of the SNRs.
Alternatively, detailed element abundance observations of SNRs can be used as a tool to identify their origins, as the abundances of some isotopes are evidence of the explosion mechanism and nucleosynthesis of the SN Ia explosions \citep[e.g.,][]{2003ApJ...593..358B, 2006ApJ...645.1373B, 2010ApJ...724L.161B, 2013ApJ...771L...9B, 2014ApJ...785L..27Y}.
The morphology and dynamics of the SNRs, on the other hand, also carry information about the surrounding environment of the SNe Ia, which could be shaped by their progenitor systems (i.e., affected by the history of mass loss of the progenitor systems) \citep[see][]{2007ApJ...662..472B, 2016ApJ...826...34Z, 2018A&A...615A.150Z}.
It may be possible that a combination of the two can shed more light on the origin of some well-known young SN Ia SNRs.

As an SN gradually fades from its peak luminosity, it starts to transit to the remnant phase, which marks the beginning of a young SNR. 
SNRs are relatively long-lived, diffuse nebular objects which are products of the interaction between the SN ejecta and the surrounding interstellar medium (ISM) or circum-stellar medium (CSM) \citep{1982ApJ...258..790C}.
As the SN ejecta gradually sweep up more materials, the SNR enters various evolution stages \citep[see][]{1972ARA&A..10..129W, 1999ApJS..120..299T}.
In this study, we simulate the SNRs through the earlier stages of SNR evolution (that is, to the end of the ejecta-dominated stage, before entering the Sedov stage), as the SNRs at these stages experience significant Rayleigh-Taylor Instabilities (RTI) \citep{1973MNRAS.161...47G, 1978ApJ...224..477S, 1992ApJ...392..118C}, by which interesting morphology features may be amplified or erased.
Especially, we focus on the effects of the companion star (or non-existence of it) on the SNR.
The companion star near the exploding WD is very likely the donor of the progenitors system, and the nature of the companion is also the defining difference between the SD and DD channels.

Previous studies and simulations have shown that intact companion stars (not yet torn apart by tidal forces or merged with exploding WDs) are likely to survive the SNe Ia explosion, and they may block or divert SN ejecta \citep{2000ApJS..128..615M, 2008A&A...489..943P, 2010ApJ...715...78P, 2012ApJ...750..151P, 2012A&A...548A...2L, 2013ApJ...774...37L, 2017MNRAS.465.2060B, 2019ApJ...887...68B, 2020ApJ...898...12Z}.
The interaction between the SN ejecta and the companion star may also result in an early bump in the SN light curves \citep{2010ApJ...708.1025K, 2015MNRAS.454.1192L, 2016MNRAS.459.1781L}.
Moreover, the interaction deposits kinetic energy into the envelope of the companion star, which makes it brighter in the short term.
However, from the point of view of the SN ejecta, the ejecta-companion interaction leaves a cone-shaped cavity in the ejecta as the companion star blocks and diverts material that was coming directly at it \citep{2012ApJ...745...75G}.
This cavity in the ejecta has been shown to have a significant influence on SNR morphology \citep{2016ApJ...833...62G}, but it is also prone to being erased by RTI fingers and re-filled with surrounding materials in the long term \citep{2021ApJ...906...93F}.

Recently, \citet{2022ApJ...930...92F} have demonstrated that asymmetries in SN ejecta, which may be caused by the explosion model \citep[such as N100, see][]{2019ApJ...877..136F} and companion interaction, can morph into identifiable features on the SNR.
They have also shown that these imprints can last for a few hundred to a thousand years into the SNR phase.
In particular, they simulated the SNR of a $\mathrm{D^6}$ model \citep{2010ApJ...709L..64G, 2013ApJ...770L...8P, 2018ApJ...868...90T}, in which the companion star is a less massive white dwarf.
Under favorable viewing angles, the post-companion-interaction cavity in the ejecta evolved into a flat feature on one side of the SNR, while other parts of the SNR remain roughly spherical.
Following their methods, we carry out a study focusing on SN Ia SNRs originated from the SD channel, while also incorporating other influences from the progenitor system, such as CSM.

The remainder of this paper is organized as follows. 
We first introduce the simulation code we use and the necessary modifications we made in Section~\ref{sec:method}, along with a detailed numerical setup. 
We then summarize the basic result of our simulations in Section~\ref{sec:results}. 
We present post-analyses of our simulation results in Section~\ref{sec:posts}. 
We discuss how to connect observations with our simulations in Section~\ref{sec:discuss}. 
Section~\ref{sec:conc} concludes this paper.

\begin{figure*}
    \centering
    \includegraphics[width=1.0\columnwidth]{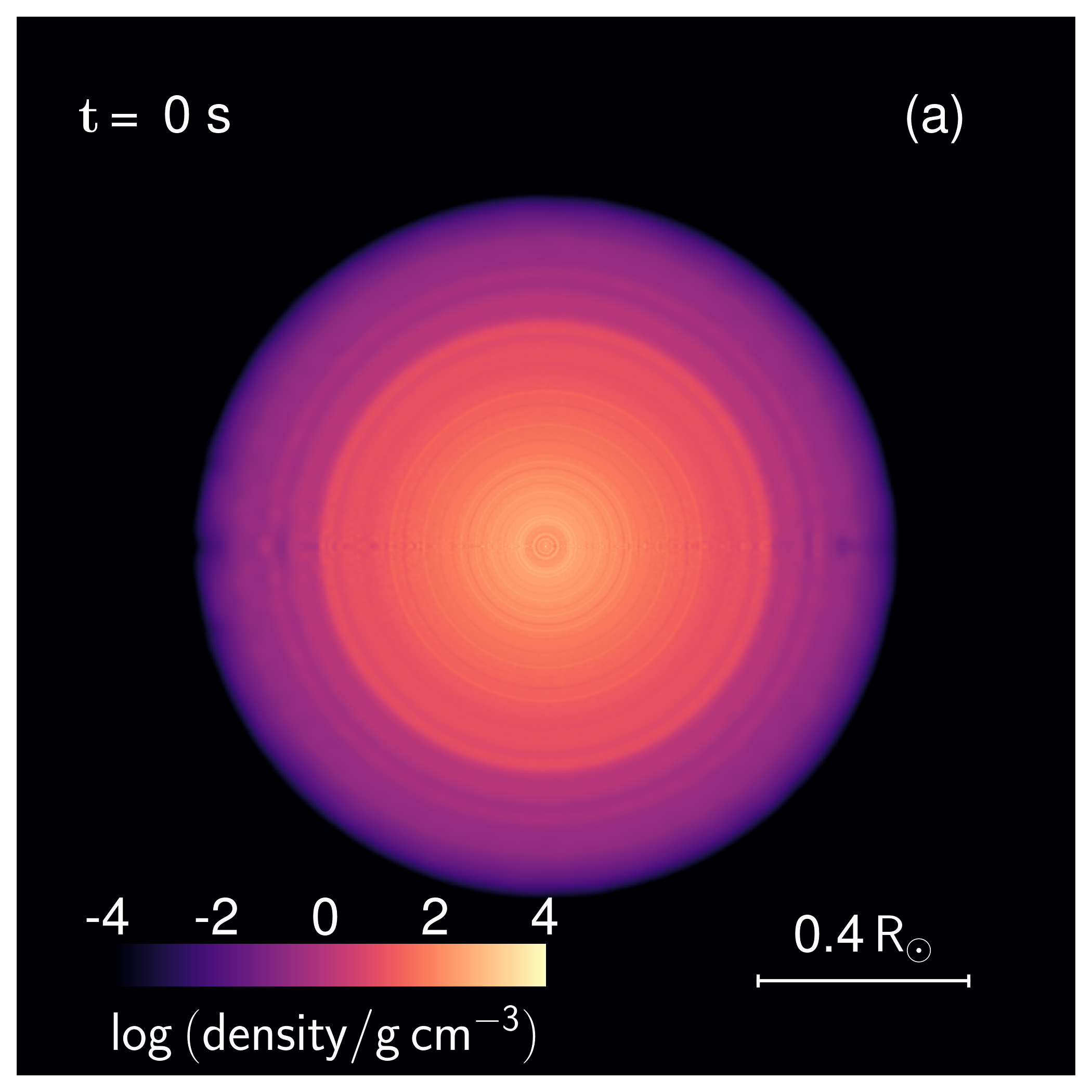}
	\includegraphics[width=1.0\columnwidth]{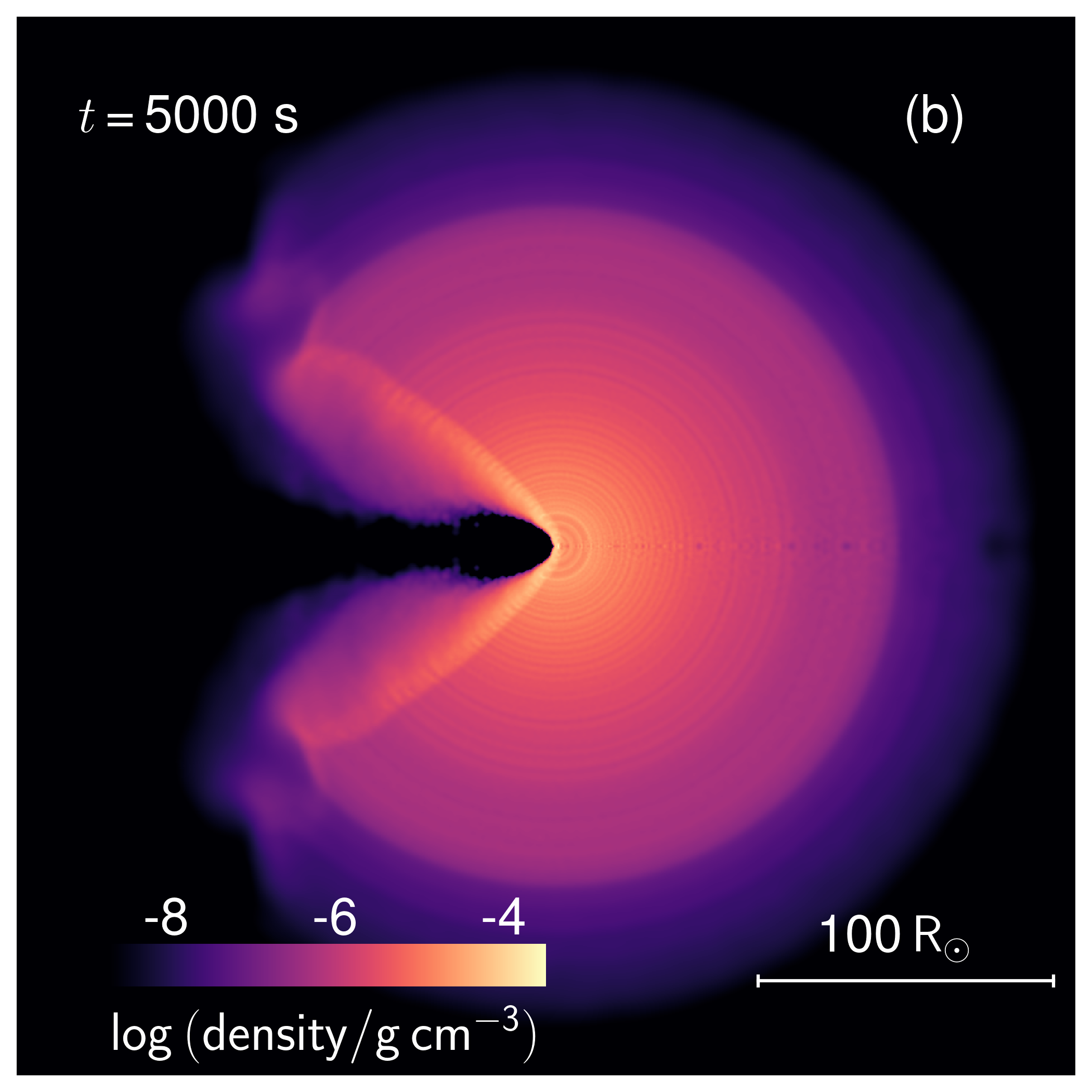}
    \caption{The SN ejecta at $\mathrm{t\,=\,0\,s}$ and $\mathrm{t\,=\,5000\,s}$, before and after interacting with the companion star.}
    \label{fig:post_impact}
\end{figure*}

\section{Methods and Numerical Setup} \label{sec:method}

\subsection{GADGET}
We used GADGET to simulate the interaction between the companion star and the SN ejecta.
GADGET is a 3D smoothed particle hydrodynamics (SPH) code that is commonly used in astrophysics simulations \citep{2005MNRAS.364.1105S}.

The progenitor system in our simulation is a binary system consisting of a Chandrasekhar-mass white dwarf and a Roche-lobe overflowing (RLOF) main sequence star (WD+MS).
The companion star was placed at the distance where it filled its Roche lobe.
We mapped the SN Ia ejecta profile of 1D deflagration model W7 \citep{1984ApJ...286..644N} into GADGET and place it at the location of the exploding WD.
We then let the ejecta expand according to its velocity profile, and it began to interact with the companion star.
At $\mathrm{t\,=\,5000\,s}$, the interaction was mostly over and the ejecta was mapped to a $\mathrm{256^3}$ grid (see Section~\ref{sec:map}), the shape and size of the ejecta are shown in Fig.~\ref{fig:post_impact}.
Detailed description of this ejecta-companion interaction simulation was described in \cite{2012A&A...548A...2L}, and their model ms22\_a was used as the input model for subsequent SNR simulations (referred to as model W7+MS in the following sections).

In the model ms22\_a, the companion star has a final mass of $\mathrm{1.21\,M_{\odot}}$, with a binary separation of $\mathrm{2.54\,R_{\odot}}$ and stellar radius of $\mathrm{0.93\,R_{\odot}}$.
During the interaction, $\mathrm{0.173\,M_{\odot}}$ of its envelop is stripped, most of these materials are moving at around $\mathrm{1000\,km\,s^{-1}}$, which are much slower than the ejecta and are expected to remain near the SNR center.
After the interaction, the companion star receives a kick velocity of $\mathrm{105\,km\,s^{-1}}$, combined with its orbital velocity ($\mathrm{237\,km\,s^{-1}}$), the final spacial velocity of this star should be $\mathrm{v_{spacial} = \sqrt{v_{orb}^2+v_{kick}^2} \,=\,259\,km\,s^{-1}}$.
This spacial velocity is insignificant compared with ejecta velocity (see Section~\ref{sec:discuss}), hence the survived companion star is likely to stay close to the SNR center during the timescale we're interested in.

\subsection{RAMSES} \label{sec:RAMSES}
Similar to \citet{2022ApJ...930...92F}, after the interaction is over ($\mathrm{t\sim5000\,seconds}$), we extracted the post-interaction ejecta and mapped them into RAMSES.
RAMSES is an open-sourced, grid-based 3D hydrodynamic code that was originally developed for cosmology simulations \citep{2002A&A...385..337T}, which also features Adaptive Mesh Refinement, self-gravitating, and magnetized fluid flows.

To accommodate the vast changes in length scale during SNR evolution ($\mathrm{<0.1\,pc}$ at t\,=\,1\,yr to $\sim$ 10\,pc at $\mathrm{t\,>\,1000yrs}$) while keeping the necessary angular resolution to resolve and trace small details in the ejecta and SNR, we employed a co-moving coordinate system, in which the computation grids expand alongside the SNR.
Similarly to \citet{2022ApJ...930...92F}, some modifications have to be made to the RAMSES code \citep{2010A&A...515A.104F}, as switching from a stationary Cartesian coordinate system to a co-moving coordinate system requires changing the hydrodynamic equations by effectively adding some source terms \citep{2007JCoPh.220..678P}.

In our SNR phase simulation, we let the SNR expand into a uniform ISM with a density of one hydrogen atom mass per cubic centimeter, or $\mathrm{\rho_{ISM}\,=\,1.0\,m_{H}/cm^{3}}$.
We assumed that the ISM has solar chemical abundance, which implies an ISM hydrogen atom number density of $\mathrm{n_{H}\,=\,0.74\,cm^{-3}}$.
Before putting the grid-based ejecta as input into RAMSES, we first scaled the ejecta to an age of $\mathrm{t\,=\,1\,yr}$ to emulate an earliest stage of free expansion.
Then we filled the non-occupied cells with ISM material, and finally transformed physical values in each cell into the co-moving coordinate system before starting the SNR simulation.
We simulated the early-stage SNR evolution from $\mathrm{t\,=\,1\,yr}$ to $\mathrm{t\,=\,750\,yr}$, when the reverse shock almost converges at the center of the remnant. Note that if the SNR evolves in a less dense environment, the timescales of the SNR evolution are longer.
The scaling law describing the relation between SN kinetic energy, SN ejecta mass, ISM density, characteristic size of SNR, and evolution timescale we use is the same as \citet{2019ApJ...877..136F} and \citet{2013MNRAS.429.3099W}.

\begin{align}
r_\mathrm{ch} &= \left( \frac{3\,M_\mathrm{ej}}{4\,\pi\,\rho_\mathrm{ISM}} \right)^{1/3} \approx 2.4\,\mathrm{pc} \\
u_\mathrm{ch} &= \left( \frac{2\,E_\mathrm{SN}}{M_\mathrm{ej}} \right)^{1/2} \approx 9{,}600\,\mathrm{km\,s^{-1}} \\
t_\mathrm{ch} &= \frac{r_\mathrm{ch}}{u_\mathrm{ch}} \approx 243\,\mathrm{yr}
\end{align}

We note that our arbitrary choice of ISM density value ($\mathrm{n_{H}\,=\,0.74\,cm^{-3}}$) is slightly higher than inferred values from some of the well observed SNRs Ia. 
But with the scaling law, we can quantitatively scale the evolution of our model to match local environments of different SNRs (i.e., SNR in denser ISM environment evolves faster and ends up smaller).
Therefore, we keep this ISM density setting throughout this work.

\begin{figure}
	\includegraphics[width=1.0\columnwidth]{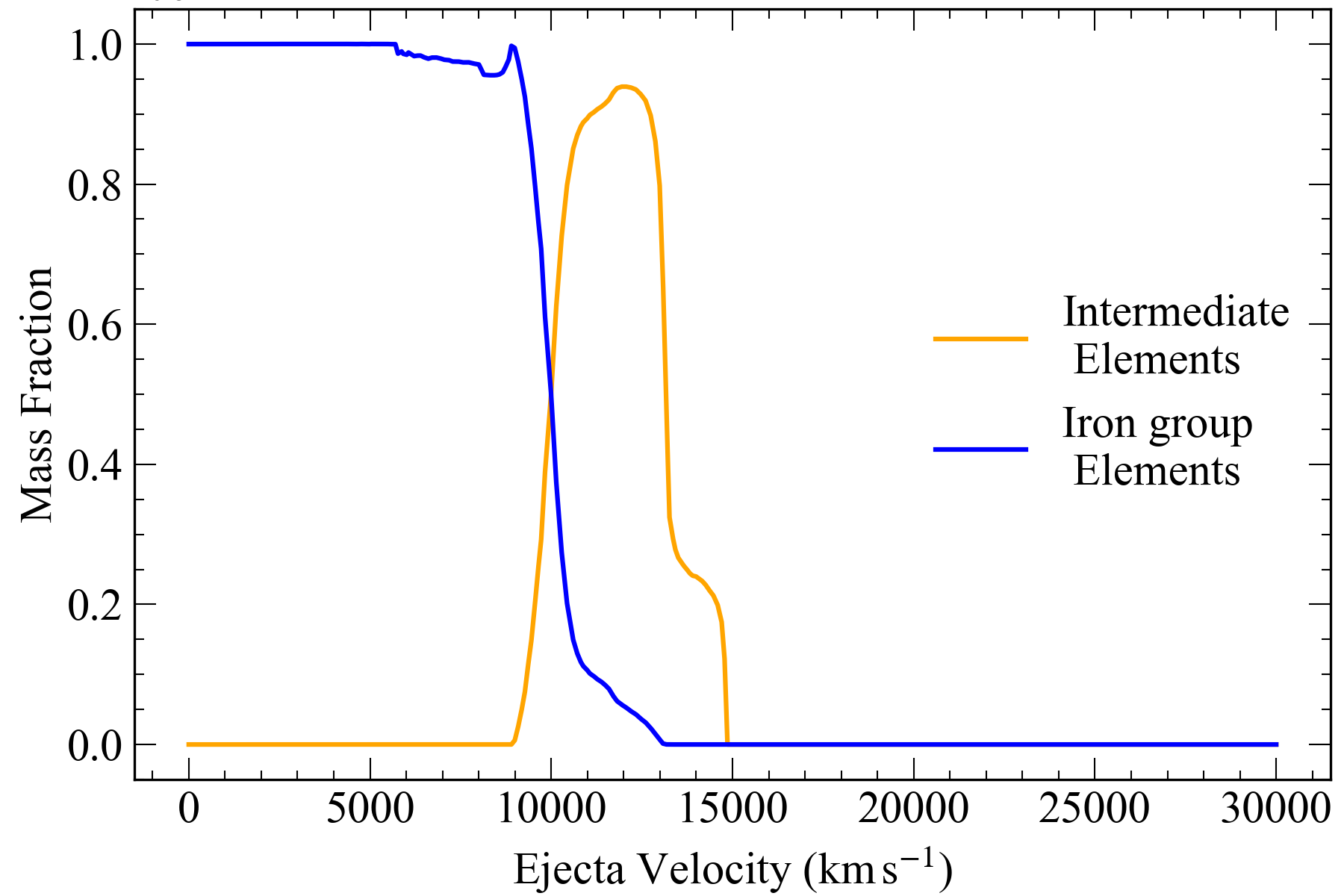}
    \caption{The element abundance profile of IMEs and IGEs in model W7.}
    \label{fig:eles}
\end{figure}

To reduce computation load and memory cost, we traced three groups of elements in our simulations.
The first group is hydrogen and helium, which denotes material that comes from ISM.
The second group is intermediate mass element (IME), which consists of S, Si, Ar, and Ca.
These elements are found in larger abundance in the outer layer of SN ejecta.
The third group is iron-group element (IGE), which consists of Sc, Ti, V, Cr, Mn, Fe, Co, and Ni.
This group also includes the radioactive isotopes of nickel and cobalt, which rapidly decay to other IGE for an SNR, hence also being included in the third group.
Each group of elements is treated as one component of the SNR and traced throughout the simulations.
The abundance profiles of IMEs and IGEs in model W7 are shown in Fig.~\ref{fig:eles}

\begin{figure*}
    \centering
	\includegraphics[width=1\columnwidth]{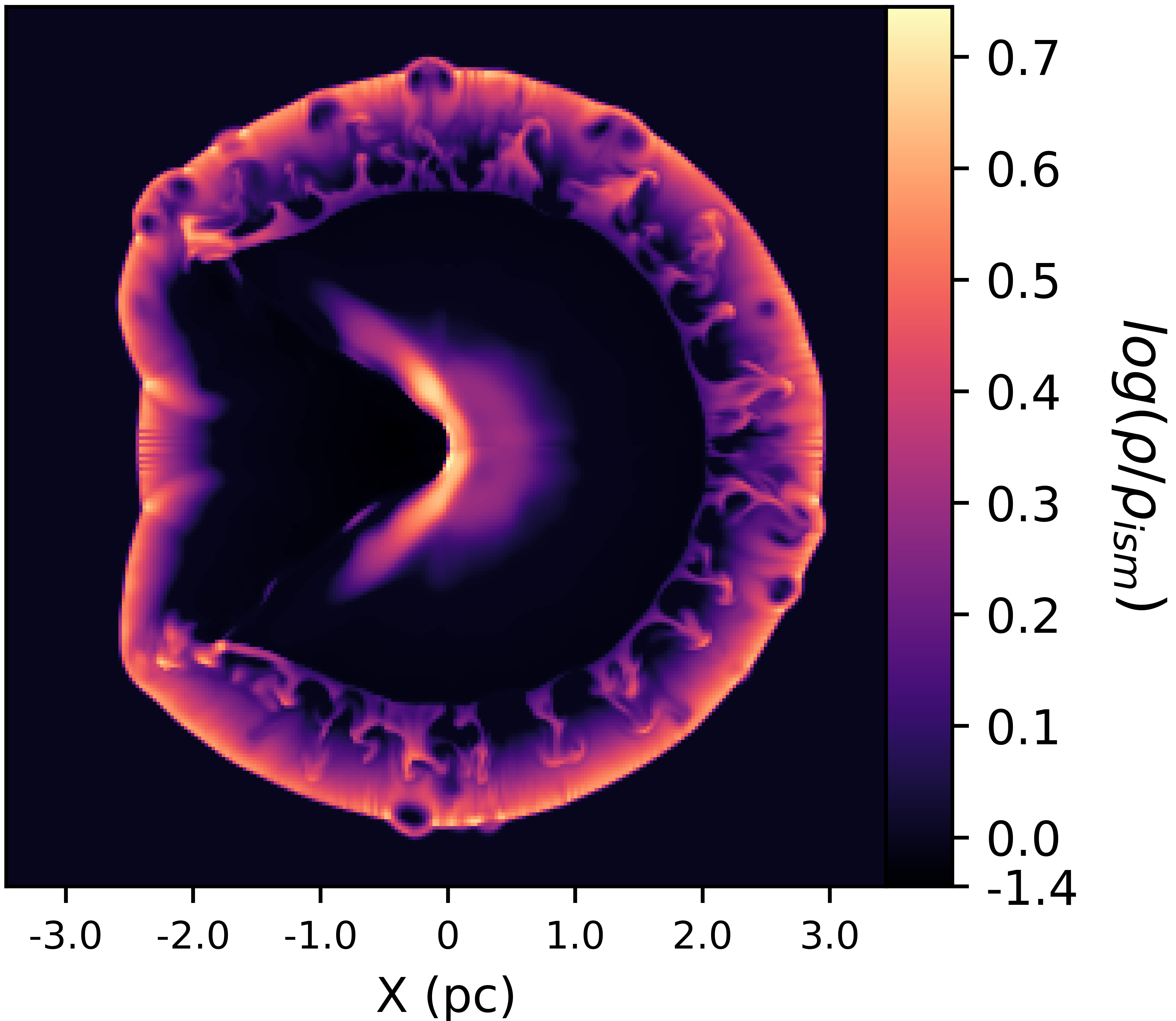}
    \includegraphics[width=1\columnwidth]{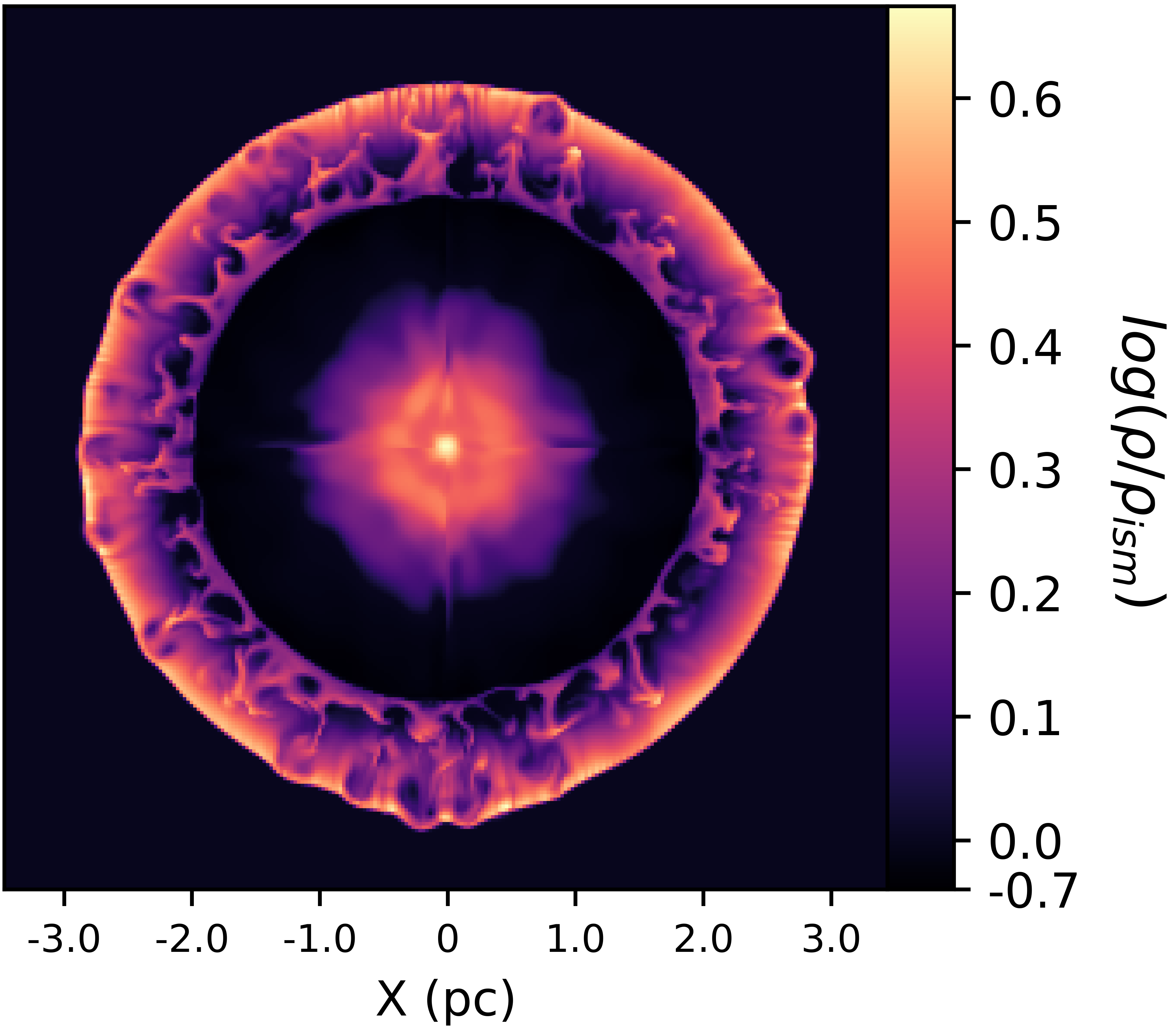}
    \caption{Left: X-Y plane 2D density slices of model W7+MS at $\mathrm{t\,=\,300\,yr}$. Right: Same but for model W7\_No\_Companion.}
    \label{fig:300Comp}
\end{figure*}

\subsection{Mapping from SPH to grid-based simulation}
\label{sec:map}
Grid-based hydrodynamic codes are traditionally better at tracing shocks and fluid-boundaries, and are more convenient for setting-up sharp density contrasts (e.g., ejecta versus ISM).
Also, grid-based codes are better at tracing hydrodynamic instabilities, such as Rayleigh-Taylor Instabilities, and turbulent structures, which are one of the main driving factors responsible for the gradual demise of the ejecta-companion interaction feature.
These are the main reasons for us to simulation result of ejecta-companion interaction from GADGET to RAMSES.
This process involved mapping particles in SPH simulation to a grid-based simulation.
Each particle from SPH represents a subset of fluid at location centered at its coordinates, and the volume affected by the particle is represented by its smoothing length, and the distribution of physical values is therefore determined by the smoothing kernel.

To map each particle from SPH to grid-based simulation, the mass and momentum of each particle was first redistributed to relevant grid cells.
The contributions from all particles were summed at each grid cell to form the density and momentum field for the grid-based simulation.
This can be described by the following equations:

For each particle, we calculated $q$, the distance between each cell and the particle center ($r_{i,x/y/z}$).

\begin{multline}
q = \frac{1}{\mathrm{SL}_i}\Bigl[
(x + 0.5 - r_{i,x})^2 
+ (y + 0.5 - r_{i,y})^2 \\
+ (z + 0.5 - r_{i,z})^2 
\Bigr]^{1/2}
\end{multline}

$i$ are the particle indices, $x$, $y$ and $z$ are the cell indices.
$SL$ is the smoothing length of each particle divided by cell length.
Then we apply cubic spline kernel $W(q)$:

\begin{align}
W =
\begin{cases}
\dfrac{1}{\pi\,\mathrm{SL}_i^3}\,W_\mathrm{cubic}(q) & \text{if } q\le 2, \\[1em]
0 & \text{otherwise}.
\end{cases}
\\
W_\mathrm{cubic}(q) =
\begin{cases}
1 - \dfrac{3}{2} q^2 + \dfrac{3}{4} q^3 & 0 \le q < 1, \\[0.5em]
\dfrac{1}{4}(2-q)^3 & 1 \le q < 2.
\end{cases}
\end{align}

Finally, we deposit the mass and momentum into the grid quantities arrays.

\begin{equation}
\rho(\mathbf{x}) \approx \sum_i m_i W(\mathbf{x} - \mathbf{r}_i, SL_i)
\label{eq:sph_density}
\end{equation}
\begin{equation}
v_\alpha(\mathbf{x}) = 
\frac{
  \sum_i \bigl( m_i / \rho_i \bigr) v_{i,\alpha}\, W(\mathbf{x} - \mathbf{r}_i, SL_i)
}{
  \sum_i \bigl( m_i / \rho_i \bigr) W(\mathbf{x} - \mathbf{r}_i, SL_i)
},
\label{eq:sph_velocity}
\end{equation}

During conversion, numerical error control was achieved by monitoring the contribution of each particle.
This produced a weight field and was later used to normalize other variable fields:

\begin{equation*}
\Delta\omega = W \cdot \frac{m_i}{\rho_i} \cdot \frac{1}{\Delta x^3}
\end{equation*}

Mapping element abundances into RAMSES is slightly more complicated due to the ejecta-companion interaction.
In order to map element abundance into RAMSES, we traced the ID of each particle in the post-interaction ejecta to find its original location in the pre-interaction ejecta.
Then the element abundances at the original location of the particle were inferred by referring to the original abundance profile of the input ejecta model (W7).
Ignoring advection of elements or any nucleosynthesis during ejecta-companion interaction, the abundance information could be then carried along with each particle to its post-impact position.

Apart from this W7+MS model, we also mapped the GADGET ejecta profile before reaching the companion star (therefore mostly spherically symmetric) into RAMSES and performed the SNR simulation, referred
to as model W7\_No\_Companion in the following sections.
This model W7\_No\_Companion represents the scenario without companion-interaction.

\section{Results} \label{sec:results}

\begin{figure*}[h!]
  \centering
  \includegraphics[width=1.4\columnwidth]{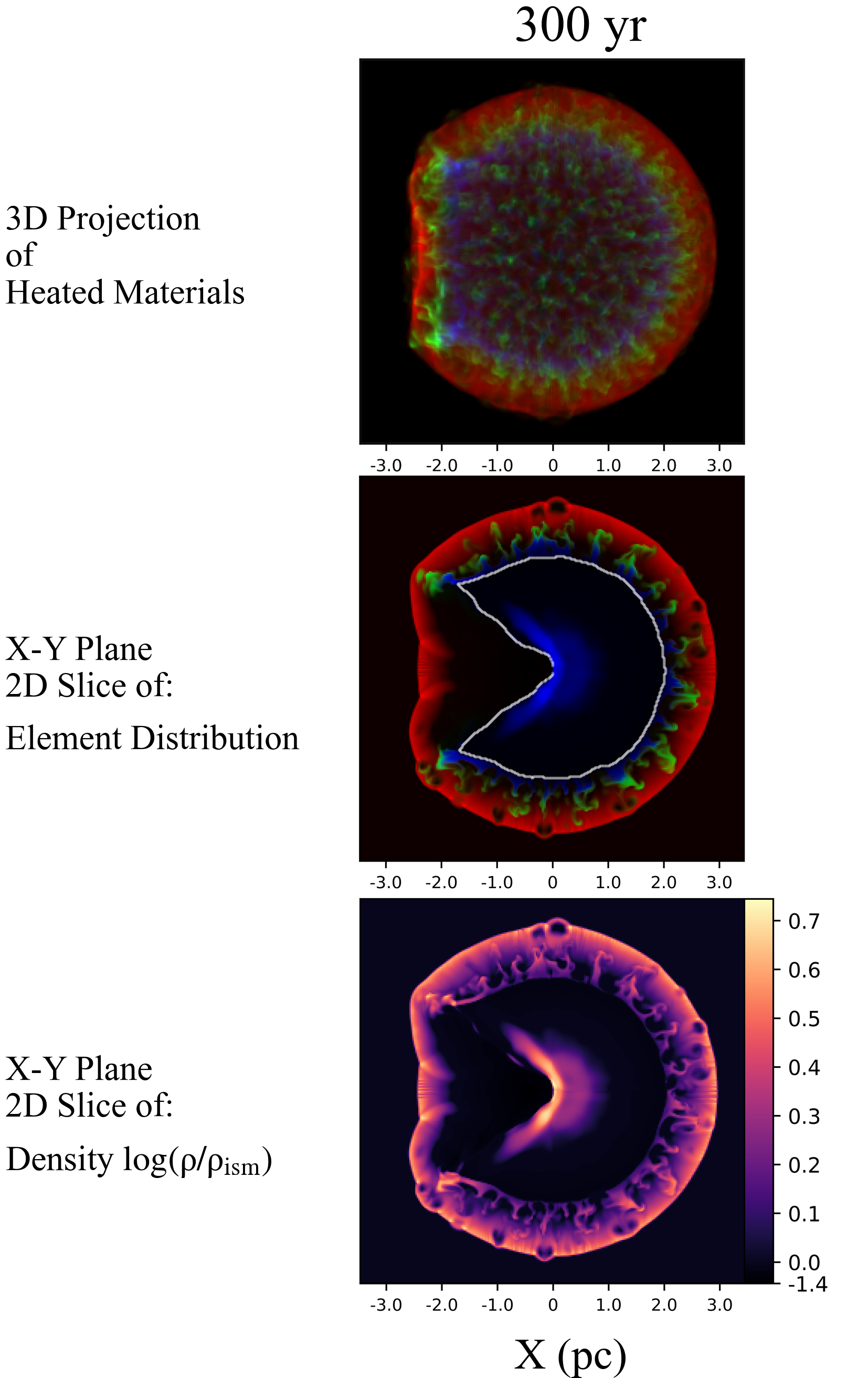} 
\caption{A Snapshot of model W7+MS at $\mathrm{t\,=\,300\,yr}$. The white lines in the middle panel mark the location of reverse shock. Red components of the top and middle panels represent H and He. Green components of the top and middle panels represent IMEs. Blue components of the top and middle panels represent IGEs. An animated version showing the whole SNR evolution sequence from $\mathrm{t\,=\,1\,yr}$ to $\mathrm{t\,=\,750\,yr}$ is available in  the ancillary files.}
\label{fig:animated_example}
\end{figure*}

A side-by-side comparison of X-Y plane 2D density slices of W7+MS with model W7\_No\_Companion at $\mathrm{t\,=\,300\,yr}$ is shown in Fig.~\ref{fig:300Comp}.
It shows that without the ejecta-companion interaction, the SNR of model W7\_No\_Companion retains a circular shape.

An animated time series of model W7+MS is shown in Fig.~\ref{fig:animated_example}. 
The top panel is the projection of shock-heated material, in which we used the square of density as a rough proxy for thermal X-ray emissivity.

The angle of projection was chosen where the imprint left over from the companion-interaction is viewed side on.
Arguably, this is a rather favorable viewing angle where the leftover feature on the SNR is most prominent.
The SNR is mostly spherical, but the cone-shaped cavity resulted from the companion-ejecta interaction forms a flat-bottom on one side of the SNR.
Similar features have also been predicted in previous studies \citep{2016ApJ...833...62G, 2022ApJ...930...92F}.
Over time, due to the effect of RTI and overpressure from surrounding ejecta over-densities, the flat-bottom feature starts to shrink and becomes less noticeable towards the end of the simulation.
Eventually, it is possible that such feature is no-longer visible or identifiable, as the morphology of older SNRs is more prone to be influenced by large-scale fluctuations of ISM (for example, the $\sim\,2000$ years old remnant of SN 185, i.e., RCW 86 as described by \citet{2011ApJ...741...96W}).

\begin{figure*}
    \centering
	\includegraphics[width=1.95\columnwidth]{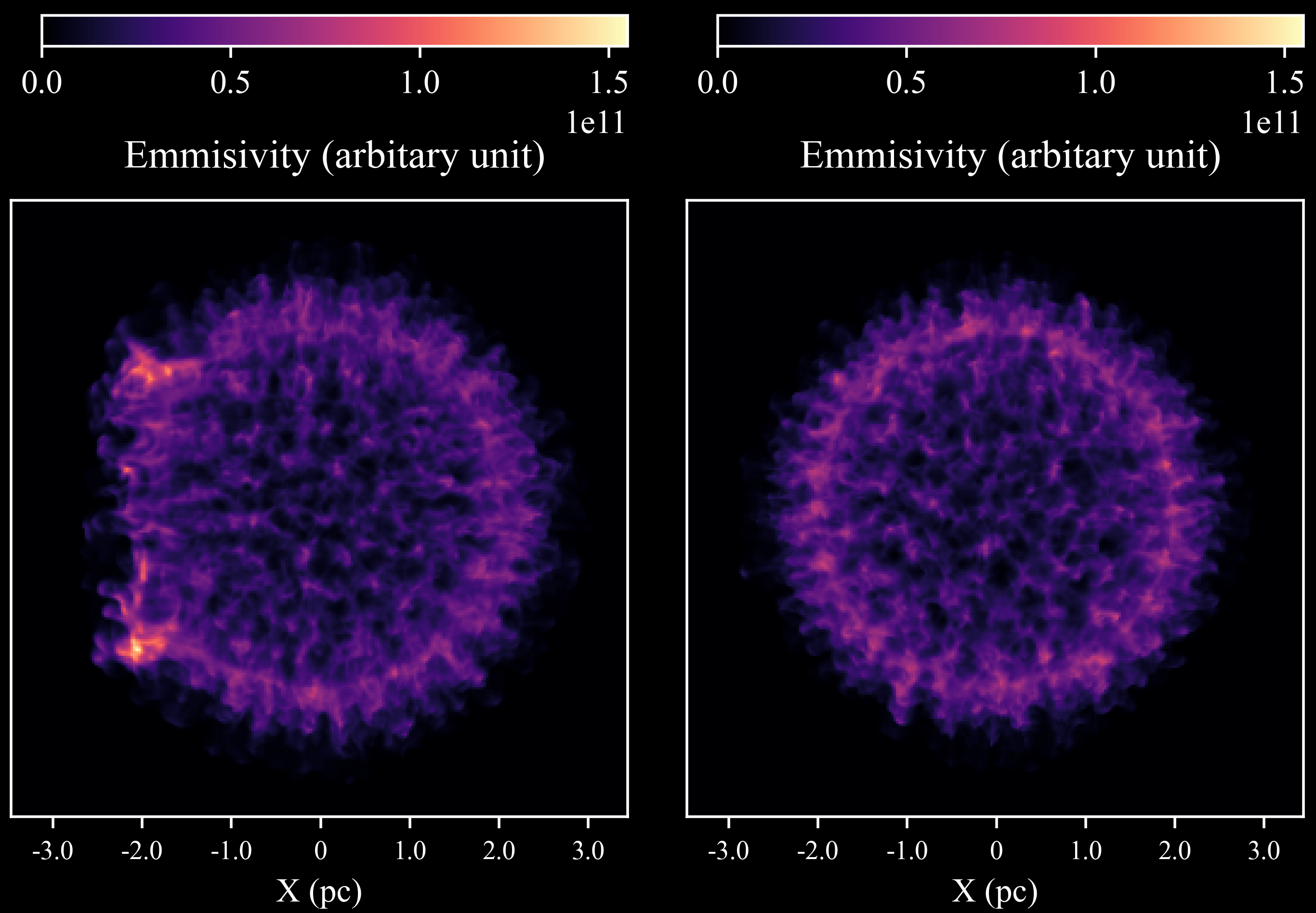}
    \caption{Left: Projection of shock-heated material density squared from model W7+MS at $\mathrm{t\,=\,300\,yr}$. Right: Same but for model W7\_No\_Companion.}
    \label{fig:proj_Comp}
\end{figure*}

We also plotted a side-by-side comparison of shock-heated material projection from the model W7+MS and W7\_No\_Companion at $\mathrm{t\,=\,300\,yr}$ (i.e. X-ray proxy), shown in Fig.~\ref{fig:proj_Comp}.
This comparison shows that without ejecta-companion interaction, the projected view of SNR also remains circular and mostly symmetric, the same as what the comparison of Fig.~\ref{fig:300Comp} shows.

\section{Post simulation analyses} \label{sec:posts}

An important driving mechanism in the ejecta-dominated phase of the SNR evolution is the growth of the RTI fingers.
These features are seen along the contact discontinuity; they disrupt the structure of the ejecta and produce small-scale features across the boundary between the SN ejecta and swept-up materials.
They are also partly responsible for the eventual demise of large-scale features such as the companion-interaction imprint, alongside with other processes (owing to inhomogeneities of ISM and overdensities of nearby ejecta materials).
Therefore, it is important to know whether our simulation successfully captures the growth of RTI fingers and their properties.

Previous studies \citep{2013MNRAS.429.3099W, 2022ApJ...940L..28P, 2023ApJ...956..130M, 2024ApJ...972...87M} have shown that power spectrum analyses can be used to check whether RTI fingers grow as expected.
Here we used the healpix package to perform power spectrum analyses of our simulation results.
We first found radial locations of contact discontinuity across the SNR and projected the radial distances onto a healpix map.
Then we extracted the power spectrum of the healpix map using the algorithm provided by the healpix package.

\begin{figure*}
    \centering
	\includegraphics[width=1.9\columnwidth]{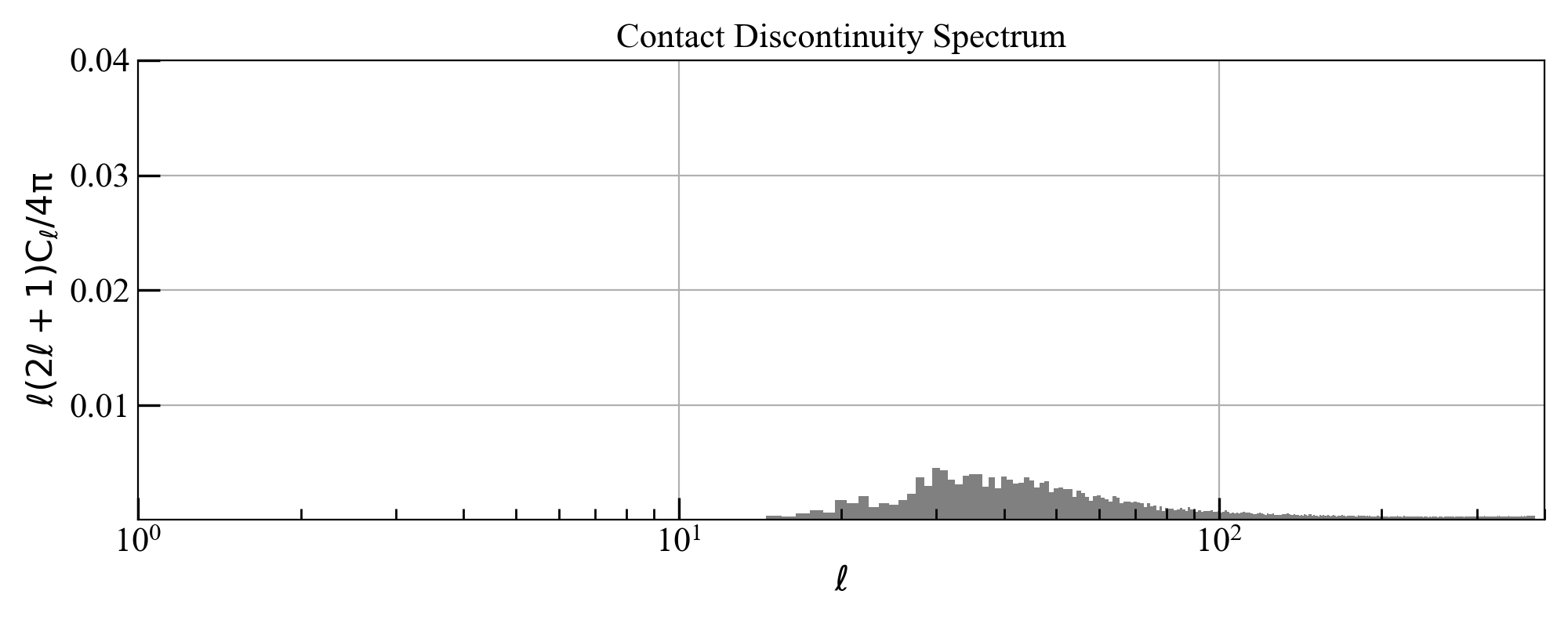}
    \includegraphics[width=1.9\columnwidth]{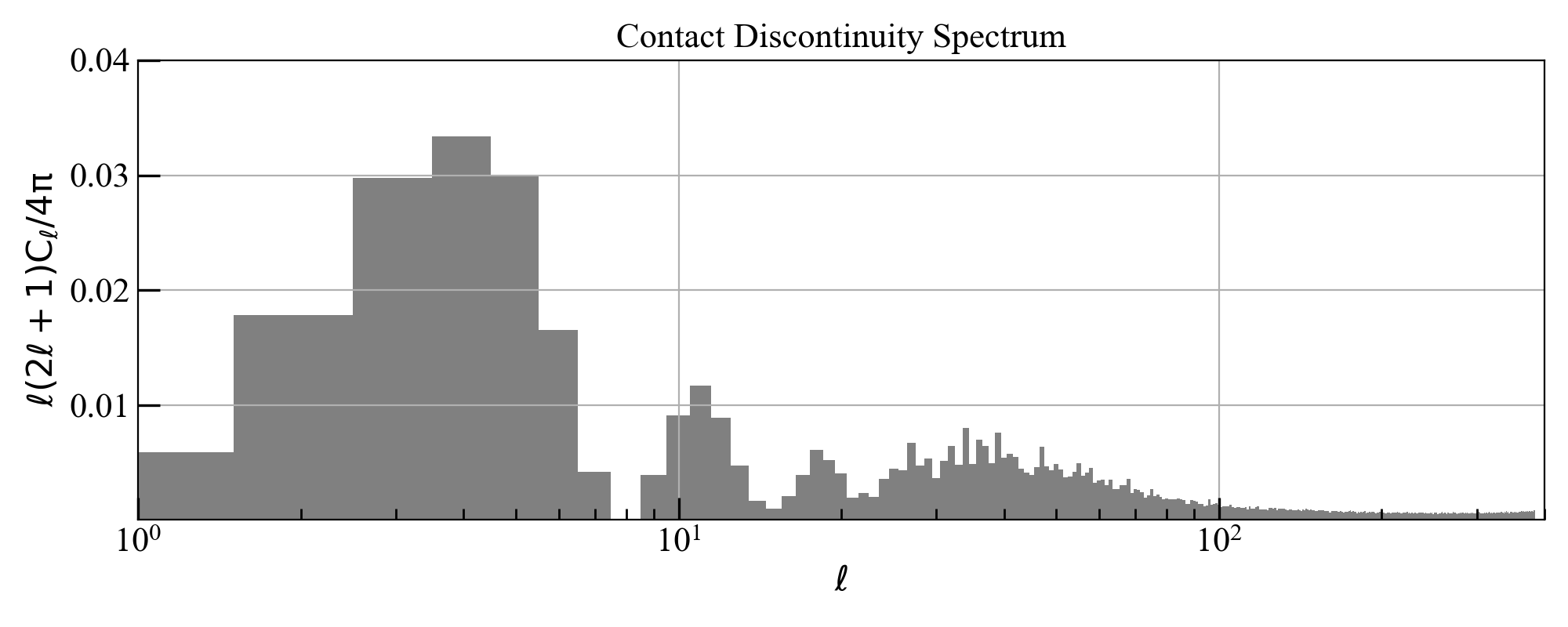}
    \caption{Upper: The extracted power spectrum of model W7\_No\_Companion. Bottom: The extracted power spectrum of model W7+MS.}
    \label{fig:power}
\end{figure*}

The extracted power spectrum of model W7\_No\_Companion is shown in the upper panel of Fig.~\ref{fig:power}, which has a peak around $\mathrm{l\,=\,40}$, similar to what has been seen in other 3D SNR simulations \citep{2019ApJ...877..136F, 2023ApJ...956..130M, 2024ApJ...972...87M}.
The power spectrum of model W7+MS (shown in the bottom panel of Fig.~\ref{fig:power}), on the other hand, shows peaks at lower frequencies (larger scales), which correspond to the features of ejecta-companion interactions.

The lowest frequency peak at $\mathrm{l\,=\,3\sim5}$ is the most prominent one, which corresponds to the impact cavity and the subsequent flat-bottom structure in the SNR.
The intermediate frequency peak around $\mathrm{l\,=\,10}$ represents the ring of over-densities caused by the ejecta material being expelled and gathered along the edge of the impact cavity.

These analyses demonstrated that our simulations successfully resolved the RTI fingers up to an appropriate angular scale where they are strongest on the power spectrum. 
It also means that any feature smaller than the dominant frequencies of RTI is very likely to be overwhelmed by the RTI and even serve as seeds for RTI growth.


\section{Comparison with observations} \label{sec:discuss}

Similarly to the $\mathrm{D^6}$ SNR simulation result \citep{2022ApJ...930...92F} and the SPH-only simulation of \citet{2016ApJ...833...62G}, our simulations showed that SNe Ia from the SD channel will leave visible ejecta-companion interaction features on the young SNR if viewed from a favorable angle.
Realistically, similar morphological features may be caused by other physical processes, such as asymmetric explosions or inhomogeneous ISM environments.
Therefore, we try to compare our simulations with observations of real-world SN Ia SNRs to see if similar features have been observed and match our model predictions.
To make this comparison, we selected a few spatially-resolved young SNRs originated from SNe Ia located in the Galaxy and the Large Magellanic Cloud (LMC). 
The sample size is heavily limited, but some of them do show interesting morphological features similar to those caused by ejecta-companion interactions in our simulations.

\subsection{SNR 0519-69.0}
SNR\,0519-69.0 has been identified as a young SNR Ia ($\mathrm{\sim\,500}$ years) in LMC by X-ray spectroscopy \citep{2010A&A...519A..11K} and light echoes \citep{2005Natur.438.1132R}.
This SNR has a prominent flat edge at its northeast (NE) side, spanning about 120 degrees, while the rest of SNR is largely circular.
At first glance, this SNR somewhat resembles the simulation result of our model W7+MS, albeit its NE flat edge has a slightly larger angular extend.
However, deeper exposures revealed some faint H-alpha and X-ray emission beyond the NE flat edge \citep[see Figure 3 of][]{2023ApJ...946...44G}, which are not predicted by SN ejecta-companion interaction models.

A greater discrepancy surfaces as inspection of its expansion velocity over a period of 10 years \citep{2022ApJ...935...78W} shows that the NE edge expands significantly slower than the rest of the SNR ($\mathrm{\sim\,2500\,km\,s^{-1}}$ versus $\mathrm{\sim\,5000\,km\,s^{-1}}$, see their Figures 2 and 3).
This contrast in expansion velocities is not consistent with our simulations, as we expect those SNR experienced ejecta-companion interaction to expand in a self-similar pattern, where each section of the SNR edge should expand with similar speed.

\begin{figure*}
    \centering
	\includegraphics[width=2\columnwidth]{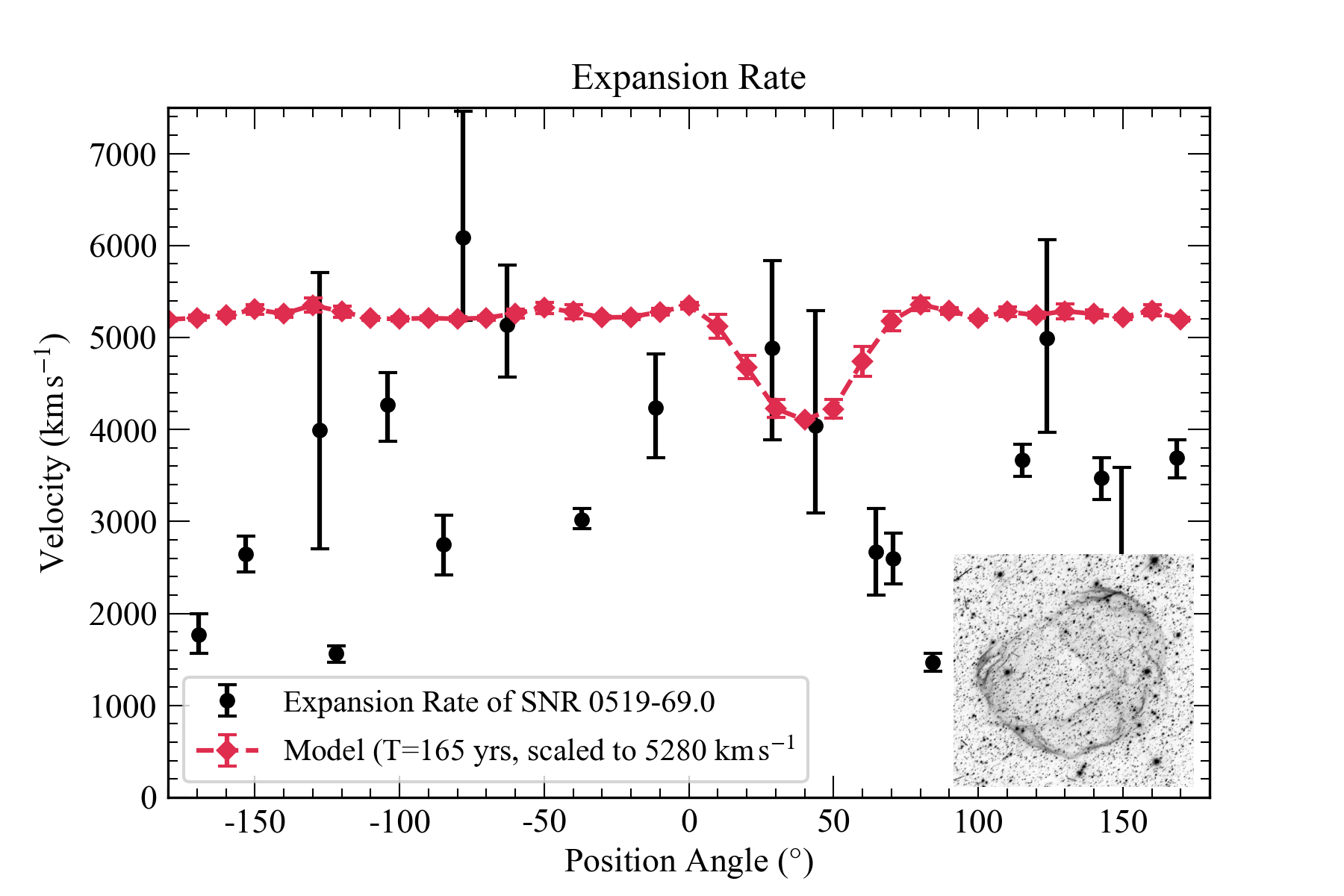}
    \caption{Front shock expansion rates of SNR\,0519-69.0 and model W7+MS. The model expansion rate shown here is an average of expansion between $\mathrm{t\,=\,160\,yrs}$ to $\mathrm{t\,=\,170\,yrs}$, rescaled to match the expansion of SNR\,0519-69.0. The HST F658N image of SNR\,0519-69.0 (visualized via Aladin) is inserted at the lower right, North is up and East to the left. The x-axis Position Angle refers on-sky orientation with regard to the SNR center, with North as zero degree and East as 90 degrees. All coordinates are in ICRS frame.}
    \label{fig:SNR0519comp}
\end{figure*}

To make a clearer comparison, we took two snapshots of our simulated model W7+MS SNR, one from $t\,=\,160\,yrs$ and another from $t\,=\,170\,yrs$, and calculated the location of the front shock in these snapshots.
To obtain a more reliable expansion measurement, we first sampled and averaged the front shock locations in 36 cone areas with radius of 5 degrees each along the X-Y equatorial plane of the modeled SNR.
The difference between each area over two snapshots represents the expansion from $t\,=\,160\,yrs$ to $t\,=\,170\,yrs$.
We then divided the measured expansion by the time duration (10 years), to get a rough estimate of the SNR expansion at $\sim t\,=\,165\,yrs$.
This process roughly mimics how real observations and measurements were made, as if our modeled SNR were viewed on the optimal angle to show the ejecta-companion interaction imprint.
Details such as a slight difference between the physical front shock and the Balmer (H-alpha) shell, projection from a 3-D shell to a 2-D ring, and how the 1-D intensity profiles across the 2-D ring translate to the exact measured locations are omitted in our rough analysis.
We adopted the same location of the SNR center as in \citet{2017ApJ...837L...7B}, and used it to derive the position angle for each measurement point of \citet{2022ApJ...935...78W}.
Finally, we scaled the front shock velocity of our model ($\mathrm{\sim\,8500\,km\,s^{-1}}$) to match that of SNR\,0519-69.0 $\mathrm{\sim\,5280\,km\,s^{-1}}$), and aligned the position angle roughly to match the model to the on-sky orientation of SNR\,0519-69.0.
The result is shown in Fig.~\ref{fig:SNR0519comp}.

\begin{figure*}
    \centering
	\includegraphics[width=2\columnwidth]{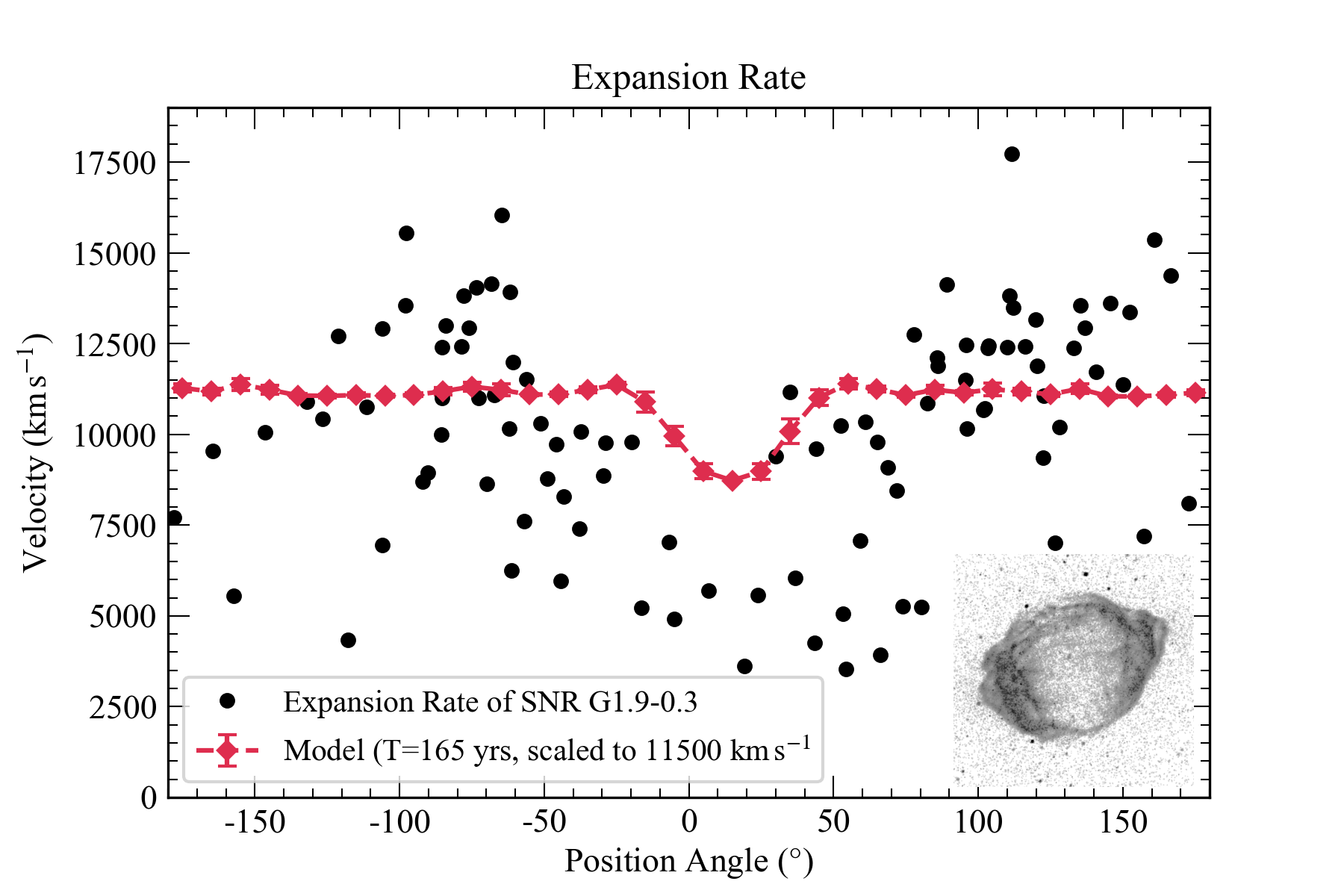}
    \caption{Front shock expansion rates of SNR\,G1.9+0.3 and model W7+MS. The model expansion rate shown here is an average of expansion between $\mathrm{t\,=\,160\,yrs}$ to $\mathrm{t\,=\,170\,yrs}$, rescaled to match the expansion of SNR\,G1.9+0.3. The Chandra X-ray image of SNR\,G1.9+0.3 (visualized via Aladin) is inserted at the lower right, North is up and East to the left. The x-axis Position Angle refers on-sky orientation with regard to the SNR center, with North as zero degree and East as 90 degrees. All coordinates are in ICRS frame.}
    \label{fig:SNRG1.9comp}
\end{figure*}

Moreover, our simulations (and other similar simulations in the literature) predict a smaller flat edge feature of about 70 degrees in size, instead of about 100 degrees seen in SNR\,0519-69.0.
The angular size of the ejecta-companion interaction imprint is strongly related to the RLOF configuration of the progenitor system.
To achieve such a large asymmetry seen in SNR\,0519-69.0, the companion star would have appeared larger in the sky of the exploding WD, hence significantly overfilling its Roche lobe, which would have been impossible unless the companion were disintegrating at the time of the SN explosion.
More violent SNe Ia explosion mechanisms (e.g. D6 and violent merger) may set up such a companion geometry, but its beyond the scope of this study as we focus on stable RLOF scenarios.
Overall, the particular shape of SNR\,0519-69.0 might be the result of on-going interactions between the SNR and surrounding CSM/ISM, and is less likely to be the result of ejecta-companion interactions.

\subsection{SNR G1.9+0.3}

SNR\,G1.9+0.3 is a very young SNR found close to the Galactic center \citep{1984Natur.312..527G, 2008ApJ...680L..41R}, and therefore heavily obscured in optical wavelengths.
Based on expansion rate measurements, the age of this SNR is about 120 years \citep{2011ApJ...737L..22C}, making it the youngest SNR in our Galaxy. 
Combined with the high expansion speed of the SNR (up to $\mathrm{13000\,km\,s^{-1}}$), this means that the fastest parts of the SN ejecta have not swept up a significant amount of ISM and therefore have not been slowed down much.
SNR\,G1.9+0.3 does show a somewhat flat structure at its northeast corner.
However, upon detailed examination of its asymmetric expansion \citep{2017ApJ...837L...7B}, it has been found that the NE part of the SNR also expands much slower than the other regions of the SNR ($\mathrm{\sim\,3600\,km\,s^{-1}}$ versus $\mathrm{\sim\,12000\,km\,s^{-1}}$).
Similarly to what we have done in the previous subsection, we made a comparison between the front shock expansion and the measured expansion of SNR\,G1.9+0.3 by \citet{2017ApJ...837L...7B} in  Fig.~\ref{fig:SNRG1.9comp}.

This disfavors ejecta-companion interaction as the only source of asymmetry in SNR\,G1.9+0.3, the interactions between the ejecta and the asymmetric CSM might be needed to explain its morphology and kinematics.
For example, a disk of material formed by previous equatorial outflow might explain the axis-symmetry features of SNR\,G1.9+0.3.
A demonstration of how localized CSM overdensities can selectively slow down parts of an SNR's front shock is provided in the Appendix~\ref{sec:apdxA}.

\subsection{SNR 1006}
SNR\,1006 originated from a historic SN Ia \citep{2002ISAA....5.....S}, and is moderately evolved in the context of our simulations.
Assuming a typical SN Ia event and taking an estimate of the ISM density of $\mathrm{\sim0.05\,cm^{-3}}$ from \citet{2007A&A...475..883A}, we found that SNR\,1006 roughly corresponds to our $\mathrm{t\,=\,400\,yrs}$ model after applying the scaling relations listed in Section~\ref{sec:RAMSES}.
The northwest quadrant of SNR\,1006 is relatively flat and bright in hard X-ray and $\mathrm{H\alpha}$.
Similarly to SNR\,0519-69.0 and SNR\,G1.9+0.3, the measured expansion speed of SNR\,1006 throughout its circumference \citep{2014ApJ...781...65W} indicates that the NW quadrant expands more slowly than other parts of the SNR ($\mathrm{\sim\,2500\,km\,s^{-1}}$ versus $\mathrm{\sim\,5000\,km\,s^{-1}}$).
This also indicates that SNR\,1006 is currently interacting or recently interacted with dense CSM near its NW edge, and the flat appearance of its NW quadrant might not be the imprint of ejecta-companion interaction.
Galactic magnetic fields could also contribute to the shape of SNR\,1006, since its symmetric axis from NW to SE is almost perpendicular to the galactic plane.

\subsection{Summary}

Overall, in this section, we tried to link the face-on ejecta-companion interaction imprint (i.e. a flat feature on one side of the SNR, caused by SN ejecta impacting the companion star and most visible when the progenitor system's orbit axis is parallel to the line-of-sight).
All three SNRs discussed in this section exhibit some morphological similarities with our ejecta-companion interaction model.
However, the detailed expansion kinematic measurements all favor recent interactions with dense CSM as the likely cause of the appearance of these SNRs.
Accounting for other SNR Ia that are more symmetric (e.g. Kepler's SNR, Tycho's SNR and SNR\,0509-67.5), we conclude that no current known young SNR Ia shows compelling morphological evidence of ejecta-companion.

However, we note that the visibility of ejecta-companion interaction imprint on the SNR is highly dependent on the viewing angle, and the current sample size of specially resolved young SNRs is insufficient to constrain the progenitor channel of SNe Ia based on the ejecta-companion interaction.
Future observations of SNRs in the local group may increase the sample size sufficiently, but the angular resolution needed for such an observation will be very challenging.
When not viewing at the best angle, the ejecta-companion interaction may still leave observable features such as over-densities or peculiar radial velocity and element distribution.
But these features may be less pronounced than the simple, flat-edge feature we try to identify in this section, and may require in-depth analysis to reveal.

\section{Conclusion} \label{sec:conc}
In this paper, we carried out simulations of interactions between a main-sequence companion star and the ejecta of exploding SN Ia, and evolved the post-impact SN ejecta to SNRs.
Similarly to previous studies \citep{2000ApJS..128..615M, 2008A&A...489..943P, 2010ApJ...715...78P, 2012ApJ...750..151P, 2012A&A...548A...2L, 2013ApJ...774...37L, 2017MNRAS.465.2060B, 2019ApJ...887...68B, 2020ApJ...898...12Z}, our SPH simulation showed that the companion-ejecta interaction leaves a cone-shaped cavity on the post-impact SN ejecta.
The subsequent SNR simulation showed that the cavity evolves into a prominent feature of the SNR during the early phases of the SNR evolution.
At favorable viewing angles, this feature should remain visible on young SNR as a flat-bottom on one side of the SNR for at least a few centuries.
This result was similar to previous studies \citep{2012ApJ...745...75G, 2016ApJ...833...62G, 2022ApJ...930...92F} but with different setups (i.e. choosing a main-sequence star as the companion star and using the W7 explosion model).
Notably, while our study focused on the single-degenerate scenario with a compact main-sequence companion, \cite{2025ApJ...982...60P} recently investigated the ejecta wakes produced by companion interaction in the double-degenerate scenario with different code (Athena++).
They also found that the companion interaction leads to structures in the supernova remnant that remain observable in X-ray emission for thousands of years.

We tried to compare observations of some known SN Ia SNRs with our simulation, but none of them showed compelling signs of a flat-bottom structure formed by the ejecta-companion interaction that matches features predicted by our model.
Although this comparison required several manipulations (such as re-scaling of shock velocities) to reconcile the simulation outputs with the observations.
These adjustments were sufficient for the main goals of the present work, as the most outstanding discrepancy between our model and the observations was the difference in the expansion velocity contrast between the flat-bottom regions and normal regions.

In the future, a deeper analysis may still provide evidence of the configuration of the progenitor system (exploded with a companion star or not) of some SN Ia SNRs.
Combined with other features, such as elements specific to the SD or DD channel explosion mechanisms, we hope that our study can help distinguish formation channels responsible for known SN Ia SNRs and eventually improve our understanding of SNe Ia.

The present study has several limitations that should be kept in mind when interpreting the results.
First, we employed a spherically symmetric SN Ia explosion model (W7); three-dimensional explosion models predict large-scale asymmetries in the velocity field (e.g., +Z / -Z half-plane anisotropies in the gravitationally confined detonation model) that may affect the remnant morphology \citep{2016A&A...592A..57S}.
Second, because we considered only the single-degenerate scenario with a compact main-sequence companion, we safely neglected the material stripped from the donor; in other WD + donor configurations, a non-negligible amount of stripped envelope material can reside within the SNR and change its morphology.
Third, the simulations assumed a CSM/ISM, whereas recent mass-transfer modeling indicates that WD accretion can produce structured, asymmetric CSM that may influence SNR morphology \citep{2016MNRAS.457..822B}.
A preliminary exploration of structured CSM (Appendix~\ref{sec:apdxA}) demonstrates that such configurations can produce strong morphology-velocity contrasts, motivating their inclusion in future work.
Finally, X-ray emissivity was approximated as the square of the density; this crude proxy reproduces observed morphologies only qualitatively, whereas realistic time-dependent temperature equilibration and non-equilibrium ionization \citep{Fujimaru2026} will be needed to connect hydrodynamic structures directly to observed line profiles.

Future work will expand the study to a broader range of progenitor channels and CSM/ISM configurations, incorporating more detailed and realistic simulations.
In particular, we plan to explore various forms of CSM in the vicinity ($\mathrm{\sim1\,pc}$) of the exploding white dwarf resulted from the mass-transfer history of the progenitor system, with the aim of reproducing the shape and velocity features of specific observed SNRs.
Long-term proper-motion observations of nearby SN Ia SNRs already indicate the presence of CSM that can produce morphological features similar to those of ejecta-companion interaction, albeit with different expansion velocity profiles; incorporating such structured CSM will therefore be a priority.
With the recent launch and commissioning of XRISM \citep{2022IJMPD..3130001T}, high spectral-resolution X-ray observations of nearby SNRs are now possible, making more realistic modeling of SNR X-ray emission essential.

\section*{Acknowledgments}
This work is supported by the National Natural Science Foundation of China (NSFC, Nos. 12288102, 12125303 , 12090040/1, 11873016), the National Key R\&D Program of China (Nos. 2021YFA1600401 and 2021YFA1600400), the Chinese Academy of Sciences (CAS), the International Centre of Supernovae (ICESUN), Yunnan Key Laboratory of Supernova Research (No. 202302AN360001), the Yunnan Fundamental Research Projects (grant Nos. 202201BC070003, 202001AW070007), the Yunnan Revitalization Talent Support Program "YunLing Scholar" project and the “Yunnan Revitalization Talent Support Program”–Science and Technology Champion Project (No. 202305AB350003).

%

\vspace{5mm}





\appendix

\section{CSM Test} \label{sec:apdxA}

\begin{figure*}
    \centering
	\includegraphics[width=1\columnwidth]{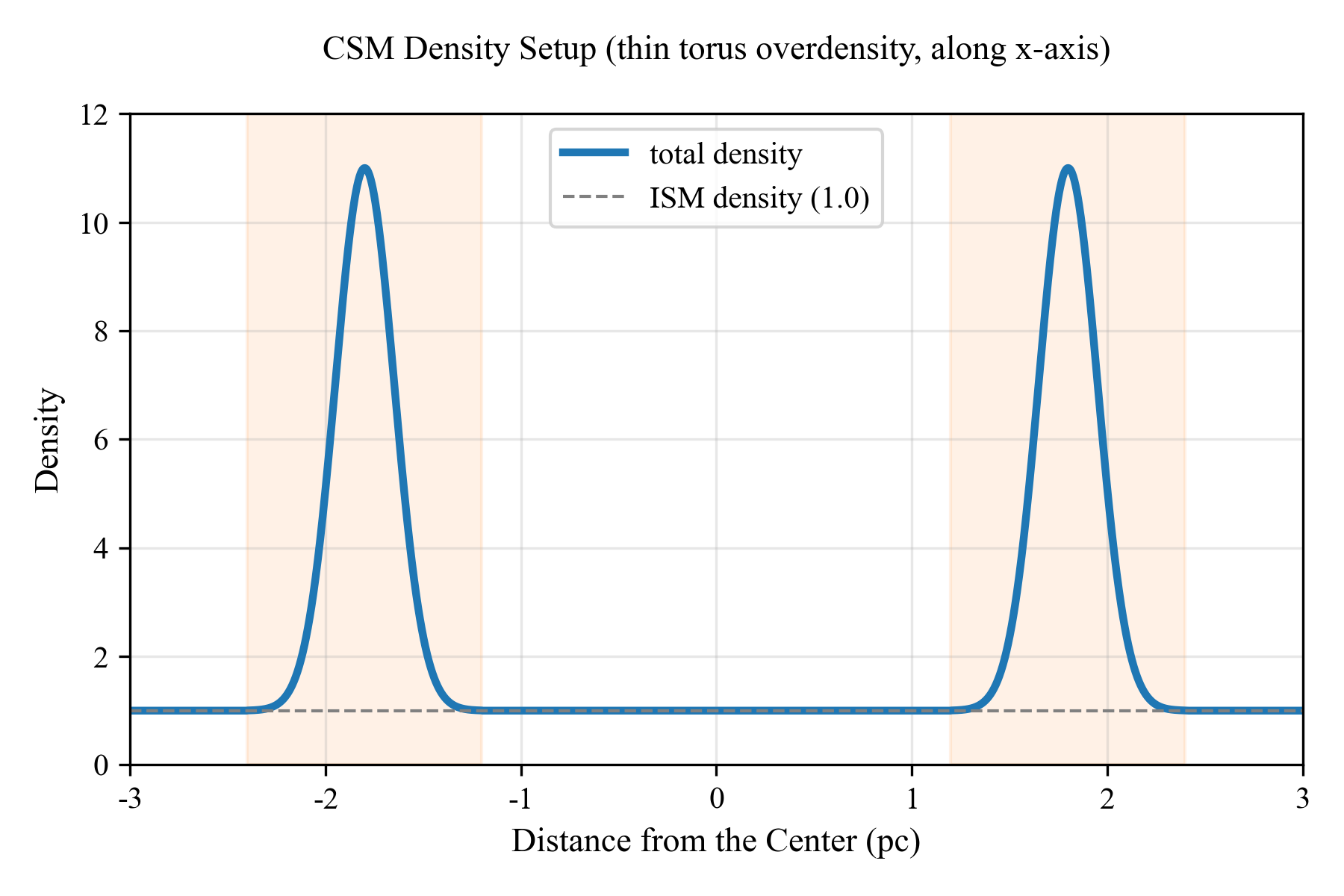}
    \includegraphics[width=1\columnwidth]{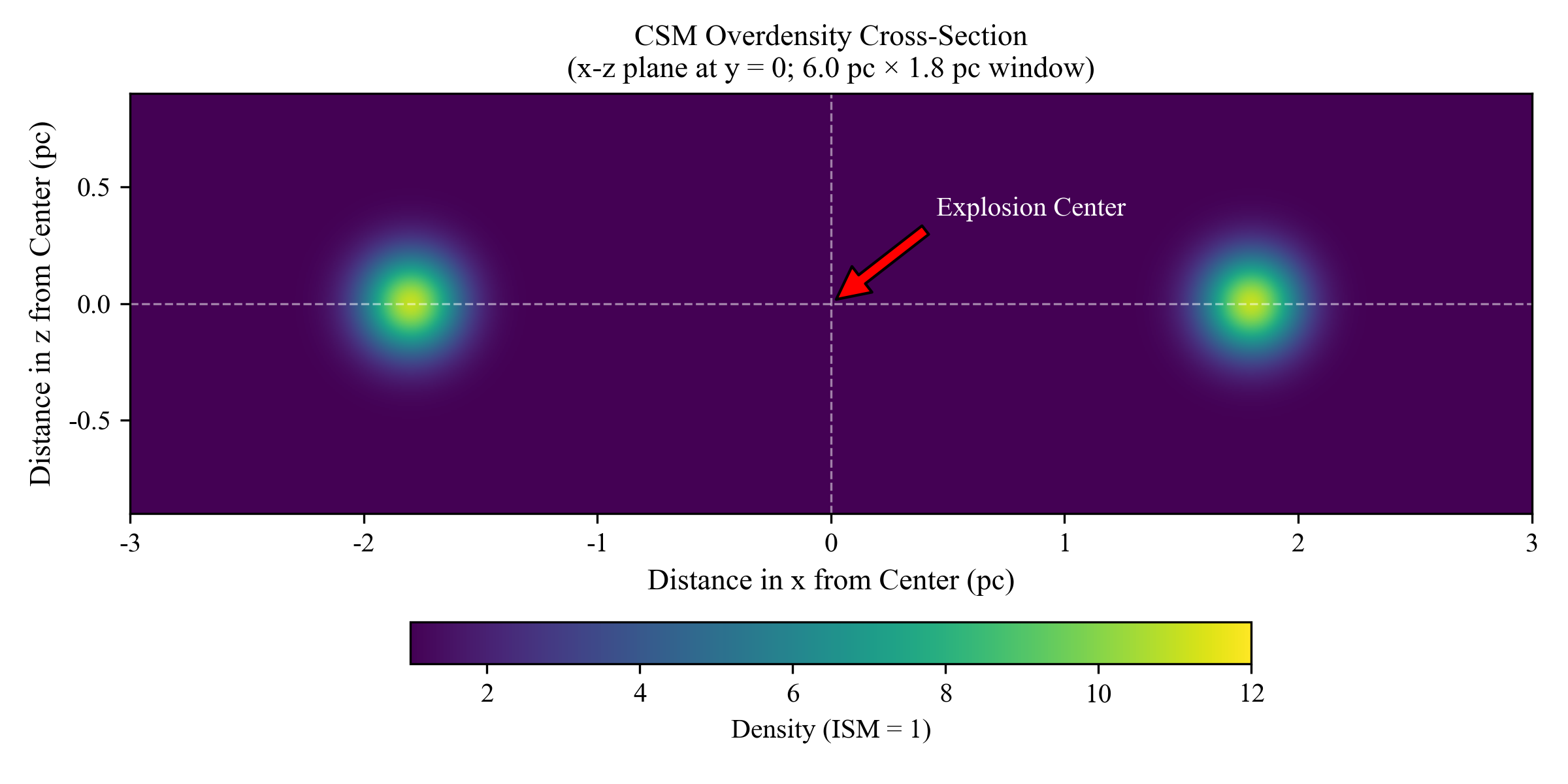}
    \caption{Density of our toy CSM setup. Upper: Density variation along the x-axis. Bottom: 2D Cross-Section of CSM density at the x-z plane.}
    \label{fig:appendix_setup}
\end{figure*}

In Section~\ref{sec:discuss} we mentioned that recent or on-going interactions between SN ejecta and over-densities in surrounding materials may cause the contrast among observed expansion velocities in different parts of the SNRs.
We also briefly discussed future plans such as incorporating CSM into our simulations to reflect certain mass-loss episodes experienced by the SN progenitor system.
Here we present a short preview of how CSM influences the structure, especially the expansion velocity of a SNR.
It also acts as a demonstration of our ability of editing surrounding environments of our SNR simulation, which our future works involving complex CSM structures will be based upon.

We placed our CSM toy model around our model W7\_No\_Companion, at a distance of 1.8\,pc (in the X-Y plane).
The torus itself follows a 2D-Gaussian density distribution, with a thickness of $\mathrm{\sigma\,=\,0.15\,pc}$.
Its total mass is about $\mathrm{0.4\,M_{\odot}}$, and is added in the standard ISM environment.
See Fig~\ref{fig:appendix_setup} for a visual illustration of this setup.

\begin{figure*}
    \centering
	\includegraphics[width=1\columnwidth]{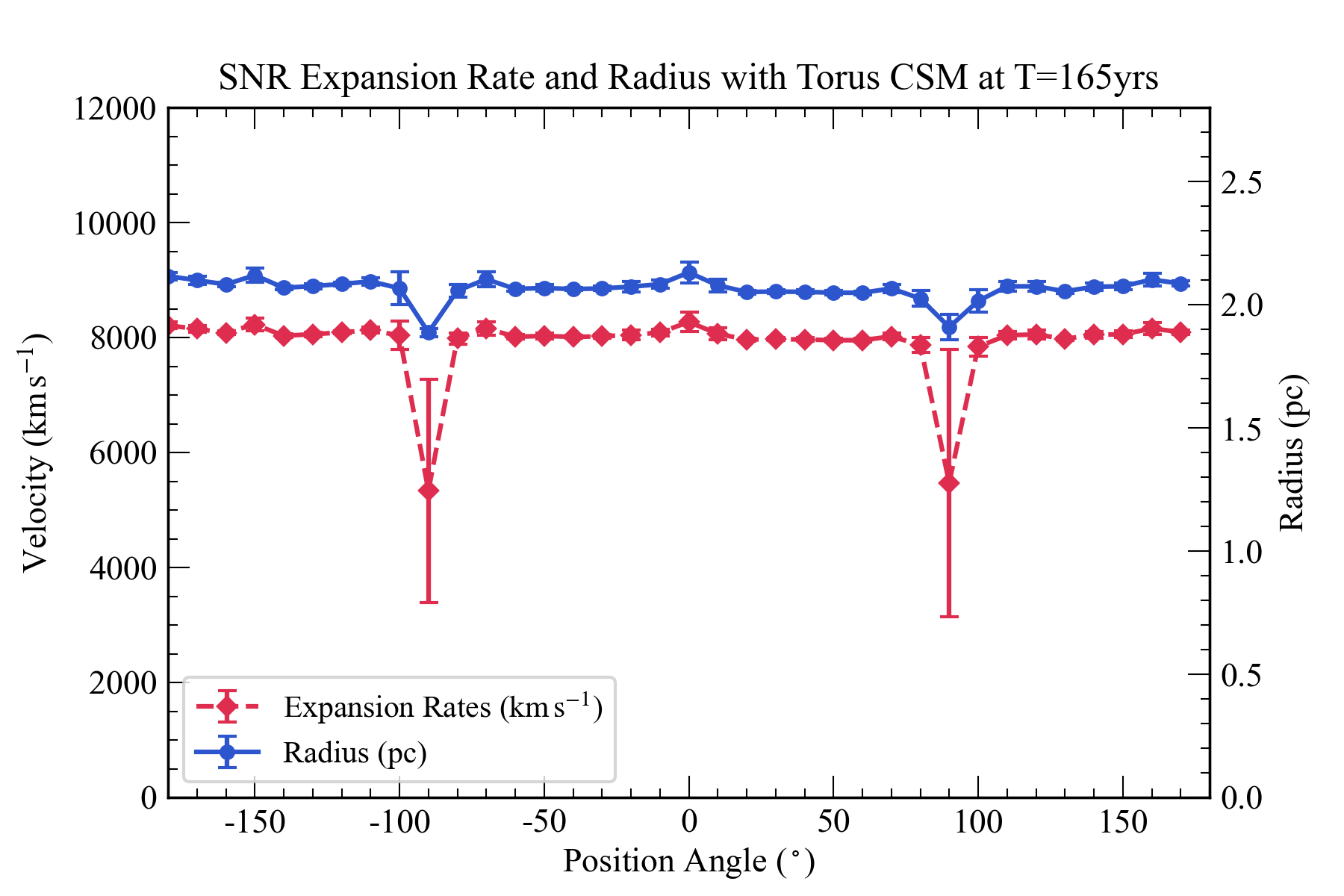}
    \caption{The red points and dashed line are the front shock expansion rates of our toy model with a torus-like CSM. The model expansion rate shown here is an average of expansion between t = 160 yrs to t = 170 yrs. The blue points and solid line are the radius of the SNR model. The two parts that affected by the CSM torus is at +90 and -90 degree.}
    \label{fig:torus}
\end{figure*}

During our SNR evolution, the SN ejecta (and some swept-up materials) encountered the CSM torus starting at $\mathrm{t\,\sim\,100\,yrs}$, which drastically slowed down the part affected by the CSM.
Similarly to what we have done in Section~\ref{sec:discuss}, we traced the expansion velocity at $t\,=\,165\,yrs$.
As shown in Fig~\ref{fig:torus}, the part of the SNR that interacted with the torus-like CSM shows subtle change (about 10 percent) in radius (1.9\,pc compared to the normal 2.1\,pc).
In contrast, the expansion velocity dropped significantly (on average 30 percent), with the slowest parts slowed to less than half of the normal expansion velocities (see the error bar).
Overall, the CSM torus in our toy model seems too thin to replicate the observed structures in SNR\,0519-69.0 and SNR\,G1.9+0.3, but it does show that recent or on-going CSM interactions can slow down the front shock significantly, while leaving the radius of the SNR relatively unchanged.

\clearpage

\bibliography{sample631}{}
\bibliographystyle{aasjournal}

\end{document}